\documentclass[dvipdfm]{pasj00}

\begin{document}
\Received{2008/07/15}
\Accepted{2008/10/03}
\SetRunningHead{Masui et al.}{Unresolved Soft X-ray Emission from the Galactic Disk}

\title{The Nature of Unresolved Soft X-ray Emission from the Galactic Disk}

\author{
K.  \textsc{Masui}\thanks{Present Address is Fuji Film Corporation.}, 
K. \textsc{Mitsuda}, N. Y. \textsc{Yamasaki}, Y. \textsc{Takei}, S. \textsc{Kimura}, T. \textsc{Yoshino}, 
 }
\affil{Institute of Space and Astronautical Science, JAXA, 3-1-1 Yoshinodai,    Sagamihara, 229-8510, Japan}
\and
\author{D. \textsc{McCammon}}
\affil{Department of Physics, University of Wisconsin, Madison, 1150 University Avenue, Madison, WI~53706 USA}

\maketitle

\begin{abstract}
Although about 40\% of  the soft X-ray background emission in 0.4 to 1~keV
range 
has extragalactic origins and thus is totally blocked by the Galactic 
absorption in midplane directions, 
it 
decreases at most by  about 20\% in midplane.  
Suzaku observation of  the direction, ($\ell$, $b$) = (235, 0),
showed an  {O}\textsc{vii} ${\rm K}_\alpha$  emission intensity 
comparable with that of the MBM-12 on cloud Suzaku observation, 
but revealed a narrow bump 
peaked at $\sim$ 0.9~keV.  The latter component is
partly  filling the decrease of the extragalactic 
component in midplane.  The feature can be well represented by a thin thermal emission
with a temperature of about 0.8~keV.   Because of the high pressure implied 
for spatially extended hot gas, 
the emission is likely 
a sum of unresolved faint sources.  We consider a large fraction of the emission 
originates from faint dM stars.  We constructed a model spectrum for spatially  unresolved
dM stars that consistently
explains the observed spectrum and the surface brightness. 
The model also suggests that  the emission from dM stars decreases very rapidly with increasing
$b$, and thus that it cannot compensate entirely the decrease of the extragalactic 
component  at $b \sim 2 - 10^\circ$. 
\end{abstract}

\section{Introduction}
\label{sec:intro}

The soft X-ray sky below 1~keV is spatially smooth after subtracting the local 
structures, such as Loop I.
In high Galactic latitudes,  $\sim$ 40 \% of the emission is attributed to 
emission from faint extragalactic objects, 
i.e. the Cosmic X-ray Background (CXB),  in the  ROSAT R45 band
 ( $\sim$ 0.44 -- 1.0~keV) \citep{McCammon_etal_2002}.  
 The rest of the emission is considered to consist of emission lines from 
 hot gas in the disk and halo of our galaxy \citep{McCammon_etal_2002}, and from the 
 Heliosphere  by the solar wind charge exchange (SWCX) process 
\citep{Cox_1998, Cravens_2000, Lallement_2004}.   
 A small fraction can  arise from intergalactic space.   
 In the Galactic midplane, the interstellar X-ray absorption column 
 density, $N_{\rm H}$, is  $\sim 10^{22} {\rm cm}^{-2}$  even in the anti-center 
 direction.  Therefore, the extragalactic X-ray photons below 1~keV are totally 
 blocked. Nevertheless, the R45 band X-ray surface brightness 
 decreases only by 20 \% or less from high Galactic  latitude to midplane. 
This issue has been known as the ``M band problem'' \citep{McCammon_Sanders_1990, Cox_2005}.  
The M band is  the name of a similar energy band in the 
Wisconsin and the Nagoya-Leiden rocket programs.  
Since X-ray photons below 1~keV 
can travel only about 1~kpc in the Galactic disk, there must be emission 
in the  midplane within 1~kpc 
which compensates partly the decrease of the extragalactic emission.  
 \citet{Nousek_etal_1982} and \citet{Sanders_etal_1983} suggested 
 hot gas of $\sim 3 \times 10^6$~K as the origin, while \citet{Rosner_etal_1981} 
 pointed out emission from dM stars can contribute $\sim 20$ \% of the total diffuse emission.  
\citet{Cox_2005} showed that if a significant fraction ($\gtrsim 1/2$) of 
emission originates from hot gas in the temperature range of 
$2.5 \times 10^6$ to $6.3 \times 10^6$ K, the hot gas must expand 
because of its high pressure.   He suggested young expanding 
superbubbles or supernova remnants evolving in low density 
region as candidates for the emission.

\begin{figure*}
\begin{center}
\begin{minipage}{0.9\textwidth}
\FigureFile(0.98\textwidth, 0.516\textwidth){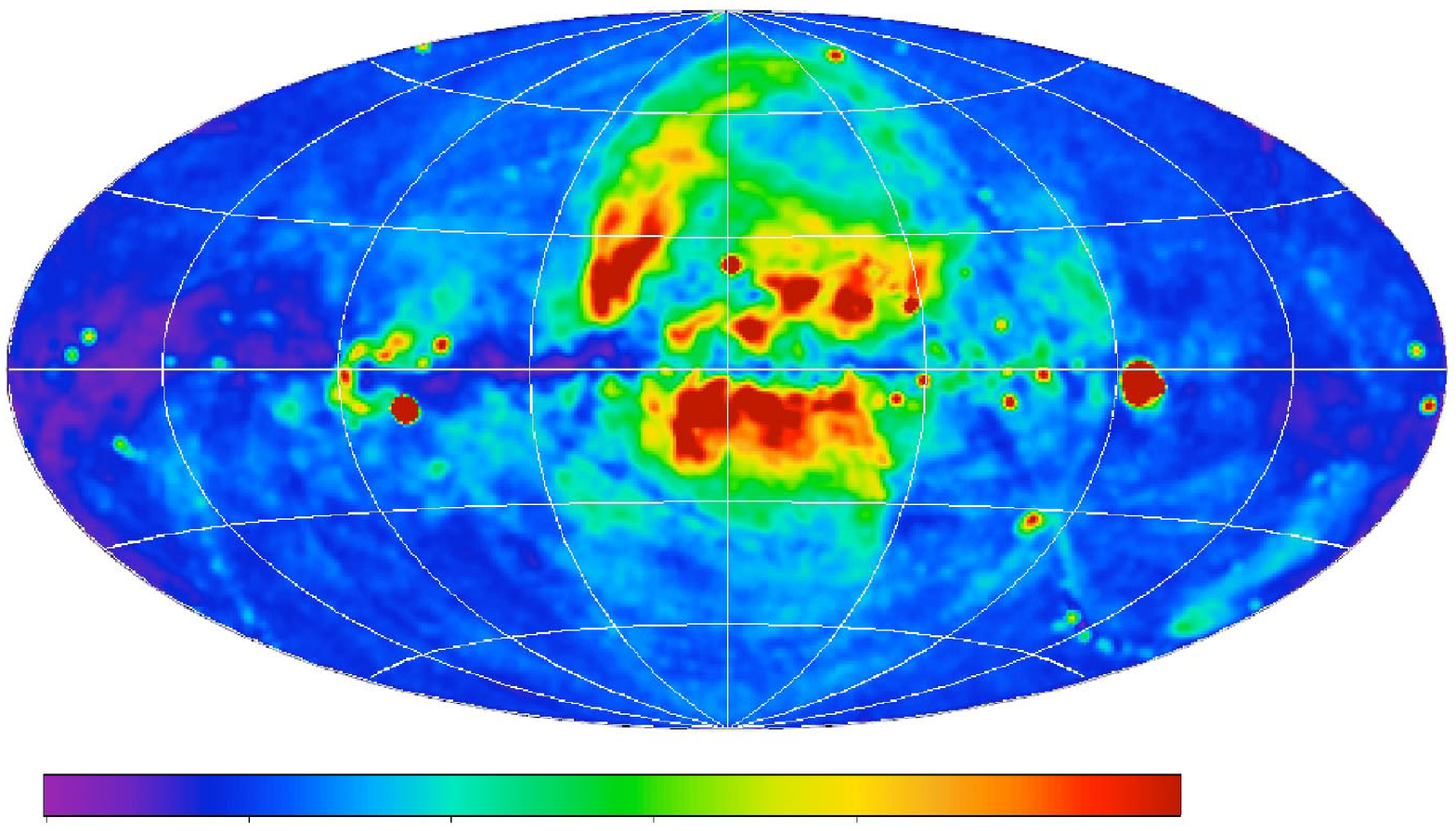}
\hspace{-0.98\textwidth}
\hspace{-1em}
\FigureFile(0.98\textwidth, 0.516\textwidth){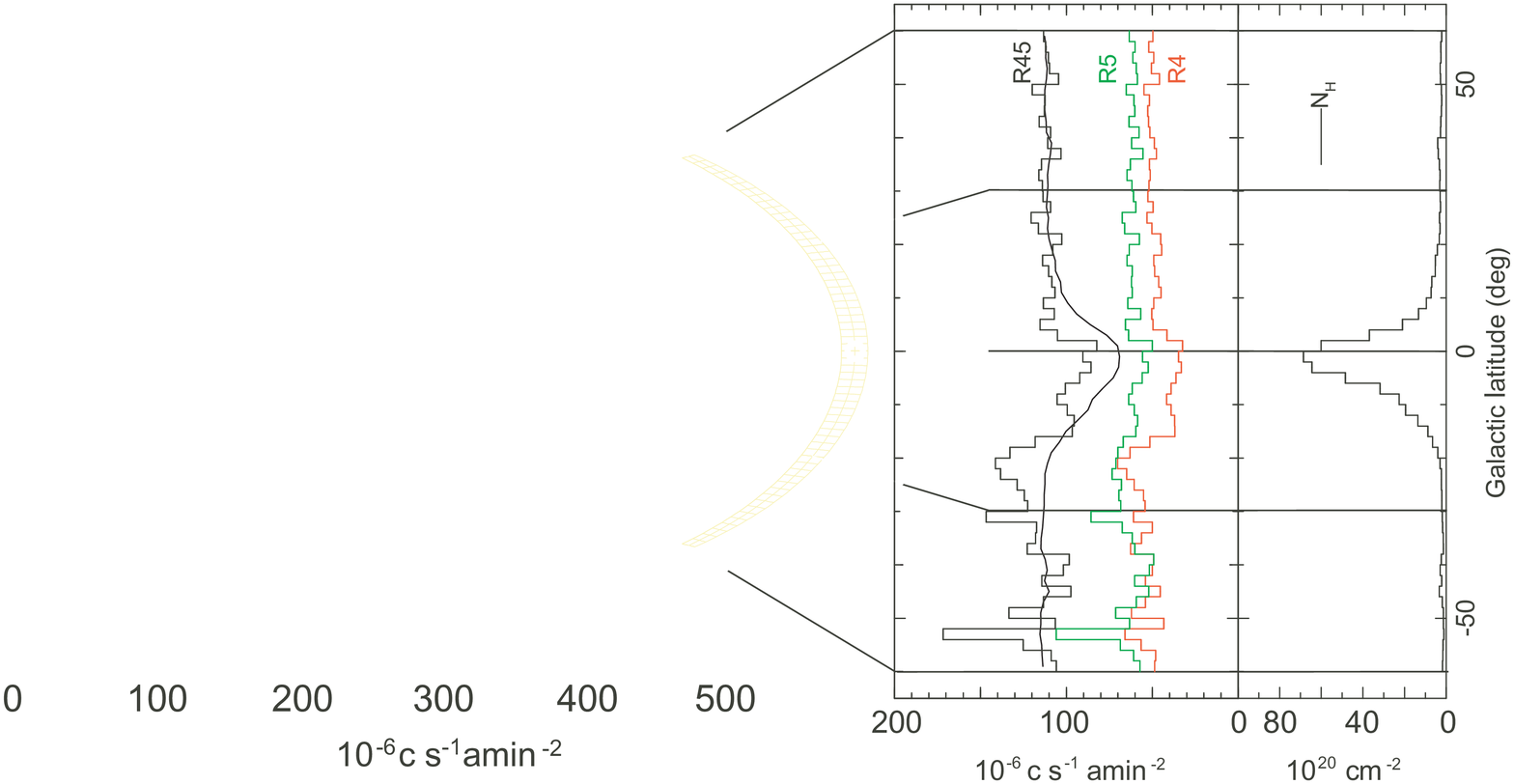}\\
\end{minipage}
\end{center}
\caption{ROSAT diffuse X-ray R45 band map \citep{Snowden_etal_1997},  
and the surface brightness and neutral hydrogen column density as functions of $b$  
along $\ell = 235^\circ$.   The thick white circle in the map indicates the pointing
direction of the present observation. 
The surface brightness and the hydrogen column density were averaged over 
rectangular areas of  size $(\Delta \ell, \Delta b) = (10^\circ, 2^\circ)$.  
The surface brightnesses in R4, R5, and R45 bands are plotted as step functions.  
A model surface brightness for R45 band which consists of
an unabsorbed constant emission  ($65 \times 10^{-6} {\rm c~s}^{-1}{\rm amin}^{-2}$) 
and the cosmic X-ray background emission ($10~ (E/{\rm 1~keV})^{-1.4}~ {\rm photons~s}^{-1}{\rm ~cm}^{-2}{\rm ~str}^{-1}~ {\rm~keV}^{-1}$  )
absorbed by the average column density is shown with a thick curve.   
There exists about $20 \times 10^{-6} {\rm c~s}^{-1}{\rm amin}^{-2}$ of excess
over the model  at midplane. The excess in 
$b = -20$ to $-30^\circ$ is partly due to one of the streaks of bright areas which meet together at the
South Ecliptic Pole.  Thus it could be due to the so-called long term enhancement \citep{Snowden_etal_1994} or scattered solar X-rays.  
\label{fig:R45map}}
\end{figure*}

As an example, the surface brightness averaged
over rectangular areas of    $(\Delta \ell, \Delta b) = (10^\circ, 2^\circ)$
along the line of $\ell = 235 ^\circ$ is plotted as a function of $b$
in Figure~ \ref{fig:R45map}.  
A model surface brightness profile  consisting of 
an unabsorbed constant emission
and the CXB 
absorbed by the average column density is plotted together with
the observational data.  For  $|b| \lesssim 10^\circ$, there is   
 $~ 20 \times 10^{-6} {\rm c~s}^{-1}{\rm amin}^{-2}$ of excess over the 
 model.  This corresponds to about 20\% of the total diffuse emission
 at high latitudes.
Note that the model curve is for  the minimum possible absorption: if the other 60\% of the R45 emission at high latitudes is produced by a hot halo, then this too would be largely absorbed in the plane, depending on its scale height structure.
The surface brightness profile suggests  asymmetry between the
profiles for $b > 0$ and $b <  0$, which is more pronounced in R4
band.  In this paper, however, we concentrate on the excess flux
at $b = 0$.   In section \ref{sec:discussion}, we will construct a model
which can consistently explain the excess at $b = 0$, although we
will find the model  cannot explain the excess in $b \sim 2 - 10^\circ$.

The origin of the excess midplane emission is not known yet although this 
problem has been known for more than 25 years.   
One major reason is that there has been 
no energy spectrum available in which emission 
line structures are resolved.  
The X-ray Imaging Spectrometer (XIS)  \citep{Koyama_etal_2007}
 on board Suzaku \citep{Mitsuda_etal_2007}
has a significantly improved
spectral line response function compared to previous X-ray 
CCD cameras, e.g. those onboard
{\it XMM-Newton} and {\it  Chandra},  in particular below 
1~keV.  Although the spectral resolution of the instrument is not high enough to resolve the
fine structure (triplet) of {O}\textsc{vii}${\rm K}_\alpha$ emission, emissions from different ions, 
e.g. {N}\textsc{vi},  {O}\textsc{vii}, and {O}\textsc{viii},  can be clearly resolved.
Combined with the X-ray telescope \citep{Serlemitsos_etal_2007}, 
the XIS also has high sensitivity for spatially extended emission.  
We have observed the direction ($\ell$, $b$) = (235, 0) with 
Suzaku for 160 ks.  
The direction was selected because this point is well outside 
the Galactic bulge and north polar spur, and because the 
direction is an average midplane direction without any 
special features, i.e. no bright X-sources in the XIS field of view, 
a typical neutral Hydrogen density,   and a typical counting rate
in the ROSAT all sky 
survey map.  We found that the{O}\textsc{vii} ${\rm K}_\alpha$ emission intensity 
was comparable with that of the MBM-12 on-cloud 
observation \citep{Smith_etal_2007} and that a narrow bump
peaked at $\sim$ 0.9~keV was compensating the decrease of the extragalactic 
component. 
This strong feature, presumably due to a blend of Ne-K and Fe-L lines, makes the $b=0$ spectrum qualitatively unlike empty-field spectra at other latitudes and requires plasma at higher temperatures than generally seen in Galactic diffuse emission.
In this paper we will show the observational results and discuss 
the origin of the excess emission.  We will construct a model spectrum for
spatially unresolved  faint  dM stars and show it can consistently explain the
observations. 

In this paper we concentrate on the XIS1 data.  This detector has a
much larger effective area below 1~keV than the other XIS sensors because
it employs backside illuminated CCD.
Throughout this paper, we quote single parameter errors
at the 90~\% confidence level unless otherwise specified.

\section{Analysis and results}

\begin{table}
\begin{center}
\caption{Log of observation \label{tbl:obs_log}}
\begin{tabular}{ll}
\hline\hline
Observation ID & 502021010 \\ 
Aim point $^{\rm a}$&  
(235.00,0.00)$_{\rm Galactic}$ \\
& = (113.33, -19.53)$_{\rm J2000.0}$\\
Observation start  time (UT) &  20:39:20, 22 April 2007\\
Observation end time (UT) &  10:04:24, 25 April, 2007\\
Net exposure time &  189.5ks\\
\hline
\multicolumn{2}{l}{
$^{\rm a}$~{\rlap{\parbox[t]{.95\columnwidth}{Aim point on the focal plane was the XIS nominal position.}}}
}
\end{tabular}
\end{center}
\end{table}

The midplane direction, ($\ell$, $b$) = (235, 0), was observed 
with Suzaku in the AO-2 period.  In Table~ \ref{tbl:obs_log} we show the log of the observation.
The XIS was set to normal clocking mode and the data format was either
$3 \times 3$ or $5 \times 5$.  The Spaced-raw Charge Injection (SCI) was on 
throughout the observation.  
We used version 2.0 processed Suzaku data. We first cleaned the data using
the selection criteria,  elevation from sunlit  and dark earth rim $>$ 20 deg, and
cut off rigidity $>$ 8 GV.
We checked the Oxygen column density of the sunlit atmosphere in the line of sight
of the screened data
and found it to be always below $10^{14} {\rm cm}^{-2}$,  which is the criterion for no significant neutral
O emission from Earth atmosphere \citep{Smith_etal_2007}.
We then checked the solar wind proton flux.  The spectrum below 1~keV could be
contaminated by the SWCX-induced emission from the geocorona if the solar wind
flux exceeds $4 \times 10^8~ {\rm protons~s}^{-1}~ {\rm cm}^{-2}$ \citep{Mitsuda_etal_2008}.  
The probability of contamination is high if the altitude of the magnetopause is lower 
than $\sim 10$ Earth radii ($R_{\rm E}$) \citep{Fujimoto_etal_2007}.  
Here, the magnetopause is 
defined by the lowest position along the line of sight whose geomagnetic field
is open to interplanetary space.  We found that the magnetopause is
higher than $10 R_{\rm E}$ throughout the observation.  However we removed
the time intervals in which the proton flux at 1AU exceeds 
$4 \times 10^8 {\rm~protons~s}^{-1}~ {\rm cm}^{-2}$ in order to avoid any contamination
by SWCX from the geocorona.  After these data selections, the total exposure time
reduced to 53 ks.

We then constructed an X-ray image in 0.3 to 2~keV energy range.  We detected 
two faint X-ray sources in the X-ray image and we removed circular regions 
centered on those point sources with radii of 2' and 1.5'. The
radii were determined by the intensities of the two sources.  The counts from
the point sources outside the circular regions are estimated, respectively,  to be less than 
5 \% and 3 \% of the diffuse X-ray emission in 0.3 to 1~keV energy range.

\begin{figure}
\FigureFile(0.9\columnwidth, 0.602\columnwidth){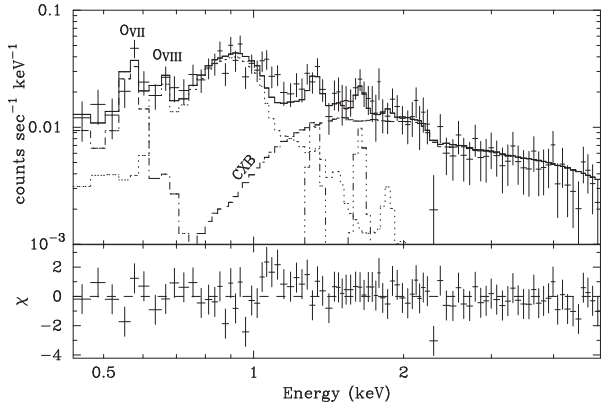}
\caption{
Observed spectrum (crosses), best-fit model and its components (step functions) of Model (A)  convolved with the instrument response function and residuals of the fit (bottom panel). The vertical error bars of data points correspond to the  $1 \sigma$ statistical errors.
\label{fig:spectrum}}
\end{figure}

The
non X-ray background  (NXB) spectrum was constructed from the dark Earth database
using the standard method in which the cutoff rigidity distributions of the on-source
and  background data were made identical \citep{Tawa_etal_2008}.   We found 
about 10 \% discrepancy 
between  the non X-ray background and the present observation 
data for XIS1 in the counting rates above 10~keV, where true X-ray
rates are expected tobe negligible, suggesting background uncertainty of this level.  
However, since in the energy range below 1~keV the non X-ray background
is only about 10 \% of the diffuse X-ray background, this level of the background 
uncertainty is negligible.

In order to perform spectral fitting, we
generated an efficiency file (arf file) for a flat field using
the xissimarfgen software version 2007-09-22
 \citep{Ishisaki_etal_2007},
assuming a 20$'$-radius flat field as the input emission of the generator.  
The degradation of low energy efficiency due to the contamination on the XIS
optical blocking filter is included in the arf file.
The pulse height redistribution matrix (rmf file) was created by 
the script xisrmfgen.

In Figure~\ref{fig:spectrum}, the energy spectrum for the accepted image region is shown.  
In the energy spectrum, one can
clearly see  a narrow bump-like  structure\footnote{We call this a narrow bump because this structure is broader than a line, but the width is only $\sim 0.3 $ keV FWHM when it is approximated with a Gaussian function} peaked at around 0.9~keV in addition to the 
{O}\textsc{vii} (0.57keV) 
emission.  
We have carefully excluded the time intervals of the observation in which the data could be contaminated by SWCX in the upper atmosphere of the Earth  (the geocoronal SWCX).  However, the SWCX in interplanetary space (the heliospheric SWCX) cannot be excluded.  The energy spectra of both the  geocoronal and heliospheric SWCX emissions reflects the abundance of ions.   \citet{Fujimoto_etal_2007} detected Fe-L and Ne-K emissions in 0.7 to 1.1 keV range in the geocoranal SWCX spectrum observed with Suzaku.  
\citet{Henley_Shelton_2008} suggested possibility of Ne-K and Mg-K emissions in SWCX based on Suzaku spectra.  However, in both cases, emissions are detected as  narrow lines with the energy resolution of the Suzaku XIS.  Thus the bump structure in the present spectrum is qualitatively very different from those SWCX spectra. 

The narrow line features at 1.3 and 1.6~keV are instrumental lines due to residuals of 
the NXB subtraction.  We first fitted the spectrum 
in the energy range 1.1 to 5~keV in order to determine the 
the intensities of the NXB residual lines. 
We employed a power-law function absorbed with a Galactic absorption for the
continuum and two narrow Gaussian functions for the NXB lines.  In the further
spectral fits, we account for these instrumental features with  
two Gaussians with all parameters fixed
at the values thus obtained.

\begin{table*}

\caption{Best fit spectral parameters}
\label{tbl:specfit}
\begin{tabular}{lccccc}
\hline\hline
Component & Model Function & Parameter (unit)  & Model (A) & Model (B) & Model (C) \\\hline
CXB & Absorption  &$N_{\rm H}$  ($10^{22}~ {\rm cm}^{-2}$) & 0.90 (fixed) & 0.90 (fixed) & 0.90 (fixed)\\
 &Power law & Photon index &   1.4 (fixed)& 1.4 (fixed)& 1.4 (fixed)\\
  &  &  Normalization$^{\rm a}$ &  $11.1 \pm 0.9$ & 11.1 (fixed) & 11.1 (fixed)\\\hline
LHB + &APEC &  k$T$  (keV) & $0.105 ^{+0.051}_{-0.032}$ & 0.105 (fixed) & 0.105 (fixed)\\
~Heliospheric SWCX &  & Normalization$^{\rm b}$ &$14.1^{+6.4}_{-9.6}$ & $13.8 \pm 3.2$ & $11.1^{+2.6}_{-3.3}$\\\hline
Broad line-like & (V)APEC &  k$T$  (keV)  & $0.766^{+0.039}_{-0.041}$ & $0.658^{+0.079}_{-0.072}$\\
~feature  & & Normalization$^{\rm b}$ &  $3.75^{+0.40}_{-0.37}$ & $2.80^{+0.45}_{-0.42}$\\
    & & O abundance (Solar$^{\rm c}$ )  & $3.1 ^{+1.3}_{-1.2}$ & 1 (fixed)\\
  &APEC & k$T$  (keV)   & &$1.50^{+0.50}_{-0.28}$\\
  & & Normalization$^{\rm b}$ &  &$3.7^{+1.7}_{-1.2}$ \\
  & Absorption &  $N_{\rm H}$ ($10^{22}~ {\rm cm}^{-2}$) & & & $0.64^{+0.27}_{-0.22}$\\
 & Bremsstrahlung &  k$T$ (keV)  & & & $0.183^{+0.074}_{-0.053}$\\
&  & Normalization$^{\rm d}$ &  &  &$4.3^{+31.0}_{-3.7}$ \\\hline
NXB residuals$^{\rm e}$&
Gaussian & Centroid (keV) & 1.32 (fixed) & 1.32 (fixed)& 1.32 (fixed)\\
 &  & Normalization$^{\rm f}$  & 0.24 (fixed)& 0.24 (fixed)& 0.24 (fixed)\\
&Gaussian & Centroid (keV) & 1.63 (fixed)& 1.63 (fixed)& 1.63 (fixed)\\
 &  & Normalization$^{\rm e}$ & 0.23 (fixed)& 0.23 (fixed)& 0.23 (fixed)\\\hline
 $\chi^2$/dof &  &                                                                                                                                                                                                                                                                                                                                                                                                                                                                                                                                                                                                                                                                                                                                                                                                                                                                                                                                                                                      &76.5/84&  61.6/85&  68.3/86\\\hline
\multicolumn{6}{l}{
$^{\rm a}$~{The unit is ${\rm photons~s}^{-1}{\rm cm}^{-2}{\rm~keV}^{-1}{\rm str}^{-1}$@1keV. }
}\\
\multicolumn{6}{l}{
$^{\rm b}$~{The emission measure integrated over the line of sight ,i.e. $(1/4\pi)\int n_{\rm e} n_{\rm H} d \ell$ in the unit of $10^{14}{\rm cm}^{-5}~{\rm str}^{-1}$. }
}\\
\multicolumn{6}{l}{
$^{\rm c}$~{Solar abundance by \citet{AG_1989}.}
}\\
\multicolumn{6}{l}{
$^{\rm d}$~{The emission measure integrated over the line of sight.
The unit is $3.02 \times 10^{-12}{\rm cm}^{-5}~{\rm str}^{-1}$. }
}\\
\multicolumn{6}{l}{
$^{\rm e}$~{\rlap{\parbox[t]{.95\textwidth}{Two intrinsically narrow Gaussians are included to represent residual instrumental emission lines.}}}
}\\
\multicolumn{6}{l}{
$^{\rm f}$~{The unit is  ${\rm photons~s}^{-1}{\rm cm}^{-2}{\rm str}^{-1}$.}
}
  
\end{tabular}

\end{table*}

We then fitted the energy spectrum in the energy range of
0.4 to 5~keV with a model consisting of three emission 
components (Model A).  
In Figure~\ref{fig:spectrum}, the best fit model function and its components convolved with the instrument
response function are shown by step functions.   
The best fit parameter values and
their statistical uncertainties are shown in the column labeled with Model (A) 
in Table~~\ref{tbl:specfit}.  
The first model component 
represents
the CXB.  
We fixed the Galactic absorption column density of this component  to
the value determined by 21 cm radio observation ($9.0 \times 10^{21} {\rm cm}^{-2}$, \cite{Dickey_Lockman_1990}).
The average AGN spectra below $\sim$ 1~keV becomes steeper than above $\sim$ 1~keV, 
and the average photon index was determined to be  1.96  below $\sim$ 1~keV by  \citet{Hasinger_etal_1993}.  
We thus first tried three different models for this component: a power-law 
function with a photon index 1.4, and broken power-law functions with 
indices of either 1.54 or 1.96 below 1.2~keV  
and with 1.4 above 1.2~keV \citep{Smith_etal_2007}.  
We set the normalization of the power-law or the broken power-law models free.  
We found all parameters of Model (A) including the normalization of the CXB at 1~keV were 
same within the statistical errors for these three CXB models.  
Therefore we will show  only the results with the single power-law function.
The best-fit value of the power-law normalization is 
consistent with typical high latitude values \citep{Revnivtsev_etal_2005}. 
The intensity of the CXB component decreases rapidly below $\sim 1.5$~keV 
because of Galactic absorption
and has negligible contribution below 1~keV. 

The second component is a  thin thermal emission model, APEC (http://hea-www.harvard.edu/APEC/), which we consider to represent the emission from hot gas in the local hot bubble (LHB) and from 
the SWCX process in the Heliosphere. 
We estimated {O}\textsc{vii} ${\rm K}_\alpha$ emission intensity 
by setting the O abundance of this APEC model to zero and substituting with a narrow
Gaussian line at the{O}\textsc{vii} ${\rm K}_\alpha$  line energy, fixing the temperature of the emission component to the best fit value. 
The{O}\textsc{vii} ${\rm K}_\alpha$ emission intensity was determined to be $2.1^{+0.8}_{-0.6}$ LU, where LU is
${\rm photons~s}^{-1}~  {\rm cm}^{-2}~  {\rm str}^{-1}$.  
This value is smaller than the{O}\textsc{vii} ${\rm K}_\alpha$ emission intensity obtained from the MBM-12 on-cloud observation by \citet{Smith_etal_2007}  with Suzaku ($3.34 \pm 0.26$ LU).  
MBM-12 is located at  60 to 280 pc from the Earth, and the{O}\textsc{vii} ${\rm K}_\alpha$ emission beyond the cloud is
completely blocked.  Thus the intensity represents the emission from the LHB and Heliospheric
SWCX.  \citet{Smith_etal_2007}  determined the emission intensity by fitting the spectrum with a power-law continuum and Gaussian lines.  However the emission intensity is dependent on how we describe the weak lines.  Furthermore,  both the  data processing and the calibration of the instruments have been improved
since the analysis by \citet{Smith_etal_2007} .
We thus re-analyzed the data and determined the emission intensity using the same
recipe used here.
We employed the same models to describe the CXB and the point source in the field of 
view as \citet{Smith_etal_2007}.  
We  found the LHB + SWCX component of the MBM-12 on-cloud direction is well described with an APEC model without absorption with k$T = 0.109 ^{+0.006}_{-0.012}$~keV, and normalization = $(13.4^{+6.1}_{-2.4}) \times 10^{14} ~{\rm cm}^{-5}~{\rm str}^{-1}$. 
The {O}\textsc{vii} ${\rm K}_\alpha$ emission was determined to be $2.93 \pm 0.45$ LU. Both the APEC model parameters and 
the {O}\textsc{vii} ${\rm K}_\alpha$ emission intensities are consistent with the present midplane
observation.

The third component is an APEC thin thermal emission model at a higher temperature
need to produce 
the  bump peaked at around 0.9~keV.  
In this model the broad feature is produced by  a sum of Fe-L and Ne-K lines.
We had to set the the O abundance of this
component free in order to better describe the {O}\textsc{viii} ${\rm K}_\alpha$ emission.
The best fit values of the temperature and the O abundance were respectively
$0.77 \pm 0.4$~keV and $3 \pm 1$ Solar.  

Although the $\chi^2$ value of the fit is statistically acceptable, 
we  find excess emission in  the 1-1.2~keV range,
which suggests  existence of even higher temperature emission.
The large required O abundance also suggests  existence of a lower-temperature
component which emits{O}\textsc{viii} ${\rm K}_\alpha$ emission more efficiently.  
Thus, the narrow bump at around 0.9~keV is  likely  multi-temperature
emission.   In the second model (Model B),  we represented the narrow bump with two
APEC components with the abundance fixed to 1.   We found the fitting is unstable
because of strong coupling among the two APEC components and the other remaining
two components.
We thus fixed the normalization of the CXB power-law function, and the  temperature of the LHB + SWCX component to the best fit values of Model (A), respectively.  In Table~ \ref{tbl:specfit}, we show the best fit parameters.  The temperatures of the emissions were
$0.66^{+0.08}_{-0.07}$ and $1.5^{+0.5}_{-0.3}$~keV, respectively.

The narrow bump at around 0.9~keV may be represented by other 
model functions.  For example, if it is a sum of emissions from faint X-ray binaries, 
it may be modeled with continuum spectra. 
Thus, as the third model,  Model (C),  we tried to fit with
a strongly absorbed continuum spectrum.  As the continuum we adopted
a bremsstrahlung model.   As shown in Table~ \ref{tbl:specfit}, the spectrum
was fitted well
with model parameters of ${\rm k}T = 0.18^{+0.07}_{-0.05}$ and
absorption column density $N_{\rm H} =  (7^{+3}_{-2}) \times 10^{21} {\rm cm}^{-2}$.

Finally, we divided the XIS image regions into two subregions and extracted the energy spectra separately.  We found that the two spectra were identical to each other and to the total spectrum within the statistical errors.  This suggests  the emission region of the narrow bump is not localized within a limited image region, but spatially extended at least to the size of the XIS field of view (18'). 
 
\section{Discussion}
\label{sec:discussion}

The pointing direction of the present observation has no  special features.  
Thus we can consider the spectrum to be representative of midplane, off-center directions. 
Therefore
a similar  narrow-bump emission is likely to fill partly the absorption of
the extragalactic component in other midplane directions, although more observations of
different midplane directions are necessary to confirm this assumption.

First we consider spatially extended hot plasma as the origin.
Assuming an average interstellar neutral  H density of 1 ${\rm cm}^{-3}$ and cosmic abundance, the absorption length of a 0.9 keV photon is estimated to be 1 kpc in midplane.   Thus 
the line of depth of the observation,  $L$,  is about 1~kpc.  Then the local H density  of the
 plasma, $n_{\rm H, hot}$, 
 is estimated from the best fit model parameters of model (A) to be 
 $n_{\rm H, hot} = 1.1 \times 10^{-3} (L/{\rm 1~kpc})^{-1/2} f^{-1/2}$, where $f$ is  the volume
 filling factor.  The pressure is then 
  $p/{\rm k} = 2.3 \times 10^4 {\rm ~cm}^{-3}{\rm ~K~} (L/{\rm 1~kpc})^{-1/2} f^{-1/2}$.
Since $f<1$, the pressure is likely to exceed the total midplane pressure derived from the 
vertical matter density and the vertical gravity, $2.2 \times 10^4 {\rm ~cm}^{-3}{\rm ~K}$ \citep{Cox_2005}.  We thus consider  
that a large fraction of the emission might arise from faint individual sources rather than diffuse gas.

The faint sources must satisfy the following three conditions: 
(1) the sources have to have energy spectra which can be approximated either by
a thin-thermal emission (k$T= 0.8$~keV) or by an absorbed ($N_{\rm H} =6 \times 10^{21} {\rm cm}^{-2}$)
bremsstrahlung (k$T =  0.2$~keV),  (2) individual sources must be faint and enough number of sources
exist in the XIS field of view (18' $\times$ 18'), and  (3) the sources must have
a relatively small vertical scale height so that the emission is enhanced only near
the Galactic plane. 

If we require at least 30 sources in the XIS field of view with a line-of-sight depth of 1~kpc, 
we need X-ray sources
with a density higher than 0.003~pc$^{-3}$.   Only normal stars and white dwarfs have high enough spatial
densities.  Among those objects, the contribution of 
main sequence M stars (dM stars) dominates the emission in the 0.3 to 1~keV energy range
because they have relatively high X-ray luminosities and
high number density \citep{Kuntz_Snowden_2001}.   Young ($\lesssim$~1~Gy) stars
have a large X-ray luminosity and dominate the total X-ray emission.  They have a small vertical
scale heights (a few 10's  pc, e.g. \cite{Bienayme_etal_1987}).

\citet{Giampapa_etal_1996} determined the energy spectra of six nearby ($< 10$ pc) dM stars
with ROSAT PSPC observations.  The spectra were found to consist of two thermal components.  
The average temperatures were k$T = 0.138$ and 0.78~keV, respectively, when they were fitted with solar-abundance plasma models.   
The two components have comparable emission measures.  
The high resolution spectrometers onboard Chandra and XMM-Newton observatories resolved the emission spectra into number of emission lines \citep{Maggio_etal_2004, Gudel_etal_2004}.  \citet{Maggio_etal_2004} determined the emission measure distribution (EMD) as a function of temperature for the dM3e star, AD Leo.   The EMD show two peaks at $\log T = 6.2$ and $\log T = 6.9$, thus at  k$T$ = 0.14 keV and 0.68 keV, which is consistent with the two temperature model from the ROSAT observations.
The temperature of the higher-temperature component is consistent with that
producing the narrow bump in the spectrum of present observation.  
The  lower temperature component of dM stars emits 
mostly {O}\textsc{vii} ${\rm K}_\alpha$ in the XIS band.  
Because opacity at {O}\textsc{vii} is about twice as high as that for 0.9~keV photons,
the line-of-sight depth contributing to the present observation is a factor of two shorter.   
Thus if we assume the narrow-bump emission arises from the 0.78~keV component of dM stars, we find the {O}\textsc{vii} ${\rm K}_\alpha$ emission intensity from the 0.14~keV component of dM star is
only about  30\%  of that of the LHB + HS-SWCX component, which is within the statistical error of the intensity. 
Thus the existence of the M-star's  lower-temperature component in the present observed
spectrum is statistically acceptable.  

\citet{Giampapa_etal_1996} also suggested that the intensity of the higher temperature component shows positive correlation with X-ray intensity during flare events.  
The EMD of AD leo obtained with the Chandra grating observation is extended 
at  temperatures above the peak at 0.68 keV up to $\sim 2$ keV. 
Van den Besselaar et al. (2003)
showed that the EMD of AD Leo is significantly enhanced in the temperature range of  k$T$ = 1-2  keV  during the decay of a flare.  
\citet{Wargelin_etal_2008} detected a flare from the nearby M star, Ross 154 with Chandra ACIS and
found that the temperature of the higher temperature component increased during the flare when it was fitted with  a two component model.  
During the decay phase of the flare, the temperature was 1.9 keV and the emission measure increased by a factor of about three compared to that during quiescence.   
The higher temperature (k$T$ = 1.5 keV)  component in Model (B) has an emission measure comparable to that of 0.7 keV component.   This could be the contribution of high temperature emission  in EMD of  dM stars in quiescence and flare.  The emission measure in 1 to 2 keV range is about 0.3 times that  in 0.7 to 0.9 keV range for AD Leo.  Then if about 20 \% of dM stars in the field of view are in flaring state, the 1.5 keV component can be explained.  Since the dM stars are highly variable (e.g. \cite{Collura_etal_1988}, \cite{Schmitt_1994}) , we consider this is plausible.

\begin{table*}
\begin{center}
\caption{Rough estimation of midplane emission measures of dM stars \label{tbl:dMstars}}
\begin{tabular}{lcccccccc}
\hline\hline
Type & Age  & $< \log L_{\rm X} >$$^{\rm a}$ & $\sigma(\log L_{\rm X})$$^{\rm a}$
        &$< {\rm EM} > $$^{\rm b}$ & Density$^{\rm c}$ & Total EM$^{\rm d}$\\
          & (Gy)  &  ($\log{\rm erg~s}^{-1}$)  &  ($\log{\rm erg~s}^{-1}$)  & ($10^{50} {\rm cm}^{-3}$) & (10$^{-3}$pc$^{-3}$) & ($10^{14} {\rm cm}^{-5}{\rm str}^{-1}$)\\\hline
M(early) & 0-0.15 & 29.19 & 0.32 & 48.3 &  4.28 & 1.72\\
              & 0.15-1 & 27.89 & 0.72 & 7.30  &  8.96 & 0.55\\
              & 1-10    & 26.86 & 0.77& 0.830 &  3.29 & 0.23\\
M(late)   & 0-0.15 & 29.12 & 0.49& 59.3  & 2.45 & 1.21\\
              & 0.15-1 &  28.2  &0.49 &  7.12  & 5.12 & 0.30\\
              & 1-10    & 27.22 & 0.49 &  0.746 &  18.8 & 0.12\\\hline
Sum      &             &            &             &                 & & 4.13\\
\hline
\multicolumn{7}{l}{
$^{\rm a}$~{Log normal probability distribution function of M star luminosity  \citep{Kuntz_Snowden_2001}.}
}\\
\multicolumn{7}{l}{
$^{\rm b}$~{\rlap{\parbox[t]{.8\textwidth}{Emission measure for the average luminosity of the log-normal distribution.  Double components of the same emission measure with temperatures of k$T=0.138$ and 0.766keV are assumed.}}}
}\\
\multicolumn{7}{l}{
$^{\rm c}$~{Average midplane star density \citep{Kuntz_Snowden_2001}.}
}\\
\multicolumn{7}{l}{
$^{\rm d}$~{Total emission measure  along a line of sight of  1~kpc length.}
}
\end{tabular}
\end{center}
\end{table*}

\begin{figure}
\FigureFile(0.9\columnwidth, 0.692\columnwidth){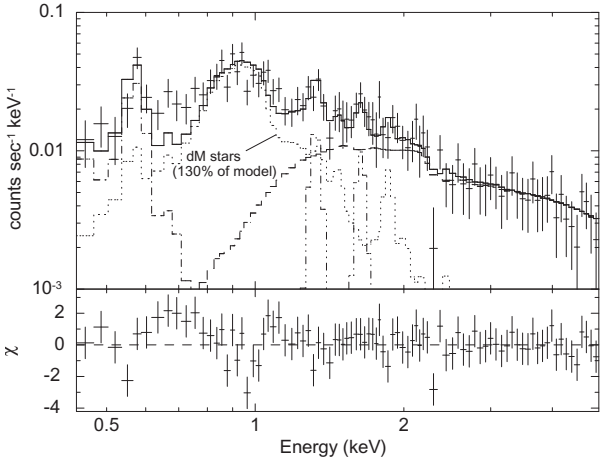}
\caption{
Observed spectrum (crosses), best-fit model and its components (step functions)  convolved with the instrument response function and residuals of the fit (bottom panel).  Here the high-temperature component of Model (A) was replaced with the model spectrum of faint dM stars. The vertical error bars of data points correspond to the  $1 \sigma$ statistical errors.
\label{fig:spectrum_mstar}}
\end{figure}

\begin{figure}
\FigureFile(0.9\columnwidth, 0.634\columnwidth){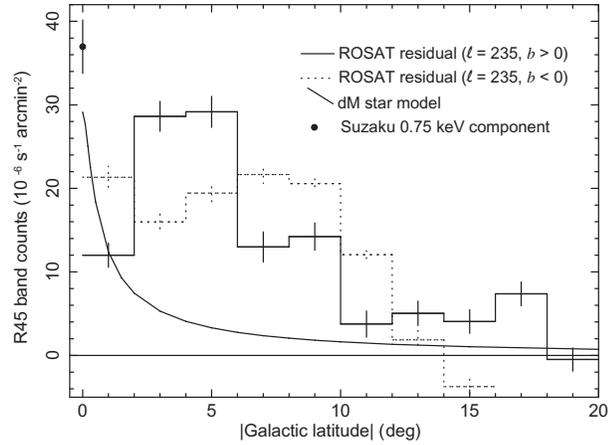}
\caption{
Residuals of R45 band surface brightness from the unabsorbed constant emission + CXB model in Figure~\ref{fig:R45map} (step functions), and expected X-ray fluxes from faint dM stars  (solid curve)  as functions of the Galactic latitude, $|b|$, along $\ell = 235^\circ$. The surface brightness of the dM star component inferred from the spectral fit to the Suzaku spectrum is shown as the filled-circle
data point at $b=0$.
\label{fig:mstar_flux}}
\end{figure}

The contributions of stars to the soft X-ray background have been estimated by several authors 
\citep{Rosner_etal_1981, Schmitt _Snowden_1990, Guillou_etal_1996, Kuntz_Snowden_2001}.  However, they estimated contributions only  for high latitudes ($b > 20^{\circ}$).  
The X-ray luminosities of stars are distributed in a relatively narrow range and the distribution function can be approximated by a log-normal function, when they are sorted by spectral type and age \citep{Schmitt _Snowden_1990, Schmitt _etal_1995} . 
Assuming a double-component thin thermal emission of k$T$ = 0.766~keV (the best fit value of Model A) and 0.138~keV with the same emission measures for the two components, we can convert the the average luminosities of the log-normal distributions to the average emission measures.  
By multiplying them with the midplane star densities and  line of sight length contributing to the diffuse emission ($\sim 1$~kpc),  and adding them together, we can roughly estimate  the total emission measure.  Using the values compiled by \citet{Kuntz_Snowden_2001} for luminosities and star densities, we obtained $4.1 \times 10^{14}~ {\rm cm}^{-5}~{\rm str}^{-1}$  (Table~\ref{tbl:dMstars}). 
This is consistent with the observed value of the emission measure, $(3.8 \pm 0.4) \times 10^{14}~ {\rm cm}^{-5}~{\rm str}^{-1}$ (Table~ \ref{tbl:specfit}).   Among different age subgroups, early and late type dM stars with ages of 0-1.5 Gy contribute a large fraction  ($\sim 70$ \%) of the emission measure.  The average number of  sources of this age group is 60 in the XIS field view.

The energy flux of the faintest class of the dM stars is estimated to be $6 \times 10^{-16}~{\rm erg~s}^{-1}~{\rm cm}^{-2}$, assuming the average luminosity of the age class, a 1 kpc distance, and the average Galactic absorption.  The sensitivity of the Chandra  Multiwavelength Plane (ChaMPlane) Survey reaches close to the limit.  However, since the ChaMPlane fields contain directions with Galactic latitude up tp $|b| = 10^\circ$, the $\log N - \log S$ relation in the 0.4-2 keV energy band is significantly contaminated by extragalactic sources \citep{Hong_etal_2005}.  Moreover, as we will show later the surface density of the dM star drops rapidly with $|b|$.  
We thus estimate the number of sources in the XIS field using the source counts in 0.4 - 2 keV range of the two low-latitude ChaMPlane fields: 
ObsID 2810 ($b = 0.18^\circ$) and  ObsID 2218 ($b = 0.14^\circ$) which have modest  exposure times: 48.8 ks and 30.2 ks, respectively.  Simply scaling the source counts of the level 3 analysis of \citet{Hong_etal_2005} by the ratio of the field of views, we obtained 70 and 24 as the expected source counts for the XIS field of view.  The nature of those sources are yet not known. However these source numbers are consistent with the above estimation of dM stars, i.e. 60.

We then constructed model energy spectra, taking into account the Galactic absorption and vertical star distribution, in order to compare the emission spectrum with observation, and to estimate the dependence of the  surface brightness on Galactic latitude.  For the neutral hydrogen column density, we used  Equations  (1) (molecular) and (2) (atomic) of \citet{Ferriere_2001}.  We scaled the midplane densities at the various galactocentric radii by the surface density in Figure~ 1 of the same paper.  For molecular gas we used the curve from \citet{Clemens_etal_1988} in the figure.  The atomic gas  surface density depends on galactocentric radius only outside 14~kpc; thus it does not affect the model spectrum constructed below. 
For the star density, we follow \citet{Kuntz_Snowden_2001}, and used Equations (A1) (0 -- 0.15 Gyr old stars)  and (A2) (0.15 -- 1 Gyr old stars) of 
\citet{Bienayme_etal_1987}.   Then the emission spectrum is given by
\begin{eqnarray}
f(E) & = & \frac{1}{4\pi} \int_{d_0}^\infty ds 
     \sum_i (\Lambda(E, T_{\rm H}) + \Lambda(E, T_{\rm L})) EM_i \rho_i(s) \nonumber \\
     & & \times \exp \left[-\sigma_{\rm abs}(E) \int_0^s ds' n_{\rm H}(s')  \right], \nonumber
\end{eqnarray}
where the integrations with $s$ and $s'$ are done along the line of sight, 
while the summation by $i$ is taken for different classes, 
i.e. spectral types and age groups, of dM stars.  
$EM_i$ and $\rho_i$ are, respectively,  the emission measure of the average luminosity star and spatial density of class $i$.  
The emission spectrum from a plasma with a temperature of $T$ and unit emission measure 
is given by $\Lambda(E, T)$, 
and $\sigma_{\rm abs}$ and $n_{\rm H}$ are respectively the X-ray absorption coefficient and the neutral hydrogen density. 
Finally, $d_0$ is the distance within which dM stars are resolved as individual stars.  We varied it from 5 pc to 20 pc and  found there was no significant difference in the resultant model spectra. 

We performed spectral fits using the model function $f(E)$ instead of a 
thin thermal emission model for the narrow bump in Model (A).  In the fit, we introduced a scaling parameter for $f(E)$ and set it free. We also set  the normalization factors of the LHB+SWCX and the CXB components free. 
The result is shown in Figure~ \ref{fig:spectrum_mstar}.  A minimum $\chi^2$ value of  
87.87 with 87 degrees of freedom was obtained for a scaling parameter value of
 $1.27_{-0.12}^{+0.10}$.  The  normalization of both the CXB component  ($10.1_{-0.8}^{+0.1.0}$)
 and the LHB+SWCX component ($13.2 \pm 3.0$) were smaller than that of Model (A), however their differences are within the statistical errors.  Therefore, both the surface brightness and spectrum of the diffuse emission at ($\ell, b$) = (235, 0) can be consistently explained by introducing the unresolved emission from dM stars.

In Figure~\ref{fig:mstar_flux}, we show the dM star model fluxes in the ROSAT R45 band  as functions of $b$ for $\ell = 235^\circ$.  We also show the residuals of the observed ROSAT R45 band flux from the unabsorbed constant emission + CXB model in Figure~~\ref{fig:R45map}.  
As shown in this figure, the dM star emissision can fill in the notch in $|b| < 2^\circ$, as well as providing the high-temperature feature seen in the Suzaku observation.  However,  there exists more  excess flux at $b \sim 2 - 10^\circ$ than the dM star model predicts.  Thus yet another emission component which should have a double-peaked latitude structure must fill the notch.  More Suzaku observations in $|b| = 2^\circ - 10^\circ$ are necessary to elucidate this unknown emission component.


\begin{thebibliography}{}
\bibitem[Anders \& Grevesse (1989)]{AG_1989}
Anders E. \& Grevesse N. 1989, Geochimica et Cosmochimica Acta 53, 197

\bibitem[Bienaym\'e, Robin \& Cr\'ez\'e (1987)]{Bienayme_etal_1987}
Bienaym\'e, O., Robin, A.C. , \& Cr\'ez\'e, M.  1987, \aap, 180, 94

\bibitem[Clemens et al. (1988)]{Clemens_etal_1988}
Clemens, D.P., Sanders, D.B., \& Scoville, N.Z. 1988, \apj, 327, 139

\bibitem[Collura et al. (1988)]{Collura_etal_1988}
Collura, A.  et al. 1988, \aap 205, 197

\bibitem[Cox (1998)]{Cox_1998}
 Cox, D.P. 1998, Lecture Notes in Physics (Berling: Springer Verlag), 506, 121

\bibitem[Cox (2005)]{Cox_2005}
 Cox, D.P. 2005, \araa, 43, 337

\bibitem[Cravens (2000)]{Cravens_2000}
 Cravens, T.E.. 2000, \apj, 532, L153

\bibitem[Dickey \& Lockman (1990)]{Dickey_Lockman_1990}
Dickey, J.M.  \& Lockman, F.J.  1990, \araa , 28, 215

\bibitem[ Ferri\'ere(2001)]{Ferriere_2001}
Ferri\'ere, K.M. 2001, Rev. Mod. Phys. 73, 1031. 

\bibitem[Fujimoto et al. (2007)]{Fujimoto_etal_2007}
Fujimoto, R.  et al. 2007, \pasj, 59, S133

\bibitem[Giampapa et al. (1996)]{Giampapa_etal_1996}
Giampapa, M.S. et al. 1996, \apj, 463, 707

\bibitem[G\"udel et al. (2004)]{Gudel_etal_2004}
G\"udel, M. et al. 2004, \aap, 416, 713

\bibitem[Guillout et al. (1996)]{Guillou_etal_1996}
Guillout, P.  et al. 1996, \aap, 316, 89

\bibitem[Hasinger et al. (1993)]{Hasinger_etal_1993}
Hasinger, G. et. al 1993, \aap, 275, 1

\bibitem[Henley \& Shelton  (2008)]{Henley_Shelton_2008}
Henley, D.B. \& Shelton, R.L.  2008, \apj 676, 335

\bibitem[Hong et al. (2005)]{Hong_etal_2005}
Hong, J. et al. 2005, \apj 645, 907

\bibitem[Ishisaki et al. (2007)]{Ishisaki_etal_2007}
Ishisaki, Y.  et al. 2007, \pasj, 59, S53

\bibitem[Koyama et al. (2007)]{Koyama_etal_2007}
Koyama, K.  et al. 2007, \pasj, 59, S23

\bibitem[Kuntz \& Snowden (2001)]{Kuntz_Snowden_2001}
Kuntz, K.D.  \& Snowden, S.L.  2001, \apj, 554, 684

\bibitem[Lallement (2004)]{Lallement_2004}
Lallement, R.  2004, \aap, 418, 143

\bibitem[Maggio et al. (2004)]{Maggio_etal_2004}
Maggio, A.  et al. 2004, \apj 613, 548

\bibitem[McCammon \& Sanders (1990)]{McCammon_Sanders_1990}
McCammon \& Sanders 1990, \araa, 28, 657

\bibitem[McCammon et al. (2002)]{McCammon_etal_2002}
McCammon, D. et al. 2002, \apj, 578, 188

\bibitem[Mitsuda et al. (2007)]{Mitsuda_etal_2007}
Mitsuda, K.  et al. 2007, \pasj, 59, S1

\bibitem[Mitsuda et al. (2008)]{Mitsuda_etal_2008}
Mitsuda, K.  et al. 2008, Prog. of  Theor.  Phys.  Suppl, 169, 79

\bibitem[Nousek et al. (1982)]{Nousek_etal_1982}
Nousek, J. A. et al. 1982, \apj, 258, 83

\bibitem[Revnivtsev et al. (2005)]{Revnivtsev_etal_2005}
Revnivtsev, M. et al. 2005, \aap 444, 381

\bibitem[Rosner et al. (1981)]{Rosner_etal_1981}
Rosner, R. et al. 1981, \apjl, 249, L5

\bibitem[Sanders et al. (1983)]{Sanders_etal_1983}
Sanders, W.T.  et al. (1983), IAU Symp. 101, 361,  ed. K Danziger, P Gorenstein,  Kluwer

\bibitem[Serlemitsos et al. (2007)]{Serlemitsos_etal_2007}
Serlemitsos, P.J.  et al. (2007), \pasj  59, S9

\bibitem[Schmitt (1994)]{Schmitt_1994}
Schimitt, J.H.M.M. 1994, \apjs 90, 735

\bibitem[Schmitt  et al.  (1995)]{Schmitt _etal_1995}
Schmitt, J.H.M.M. et al.  1995, \apj 450, 392

\bibitem[Schmitt \& Snowden (1990)]{Schmitt _Snowden_1990}
Schmitt, J.H.M.M. \& Snowden, S.L. (1990), \apj, 361, 207

 \bibitem[Smith et al. (2007)]{Smith_etal_2007}
Smith, R.  et al. 2007, \pasj,  59, S141

\bibitem[Snowden et al. (1994)]{Snowden_etal_1994}
Snowden, S.L. et al.  1994, \apj, 424, 714

 \bibitem[Snowden et al. (1997)]{Snowden_etal_1997}
Snowden, S.L.  et al. 1997, \apj,  485, 125

\bibitem[Tawa et al. (2008)]{Tawa_etal_2008}
Tawa et al.  2008, \pasj,  60, S11

\bibitem[van den Besselaar et al. (2003)]{vdBesselaar_etal_1987}
van den Besselaar et al. 2003, \aap 441, 587

\bibitem[Wargelin et al. (2008)]{Wargelin_etal_2008}
Wargelin et al.  2008, \apj,  676, 610


\end{thebibliography}
\end{document}